\documentstyle[12pt]{article}

\newcommand{\Vec}[1]{\mbox{\boldmath$ #1 $}}
\newcommand{\Unit}[1]{\hat{\Vec{ #1 }}}
\newcommand{\C}[1]{{\cal #1 }}
\newcommand{\Int}[1]{\int\!\frac{d^{3} #1}{(2\pi)^{3}}}
\newcommand{\Tr}{\mathop{\rm Tr}\nolimits}

\begin{document}
\baselineskip 24pt
\begin{titlepage}
\begin{flushright}
HUPD-9703\\
\end{flushright}

\vspace{1.5cm}

\begin{center}
{\Huge Phenomenological interaction between current quarks}\\

\vspace*{1.5cm}

{\large Michihiro HIRATA
 \footnote{e-mail address:
                  hirata@theo.phys.sci.hiroshima-u.ac.jp}
      , Naoto TSUTSUI
 \footnote{e-mail address:
                  tsutsui@theo.phys.sci.hiroshima-u.ac.jp}
 \\
 \vspace*{3mm}
        Department of Physics, Hiroshima University\\
        Higashi-Hiroshima 739, Japan}
\end{center}

\vspace*{1.5cm}

\begin{abstract}
\baselineskip 24pt
We construct a phenomenological model which describes the
dynamical chiral symmetry breaking (DCSB) of QCD vacuum and
reproduces meson spectra.  Quark condensates, the pion decay
constant, and meson spectra are well reproduced by
phenomenological interaction which consists of a linear
confining potential, a Coulombic potential, and the 't~Hooft
determinant interaction.  In this model, the 't~Hooft
determinant interaction plays a important role not to only
$\eta,\eta^{\prime}$ mass difference, but other meson masses
through DCSB.
\end{abstract}

\end{titlepage}

\section{Introduction}
It is widely accepted that QCD is the correct theory of
strong interactions.  However, it is not so easy to treat
the hadron-hadron scattering phenomena by QCD directly.
Even the lattice QCD may not be able to calculate phase
shifts of partial waves in near future, although there is an
attempt to give some hadron-hadron scattering lengths.  In
this situation it has still meaning to model QCD in a way
which has the features of QCD to describe hadron-hadron
scatterings from the quark freedom.

According to ref.\cite{glueball}, we assume the existence of
a set of phenomenological interactions $H^{I}_{phen}$ such
as the residual interaction $H^{I}_{QCD} - H^{I}_{phen}$ can
be perturbatively treated.  There exists at the present time
no model reproducing the experimental data and incorporating
simultaneously all the non-perturbative phenomena, such as
dynamical chiral symmetry breaking (DCSB) or confinement,
which are the basic aspects of QCD.  The models which
incorporate DCSB and confinement by using the instantaneous
confinement potential are at odds with phenomenology
\cite{adler}, \cite{yaouanc}, \cite{kocic}, \cite{hirata},
\cite{alkofer1}.  The covariant models which preserve the
confinement give rather good fitting for the quark
condensate and the pion decay constant \cite{alkofer2},
\cite{toki}.  However, these models must be regularized for
the infrared divergence, and furthermore in the covariant
models it is very difficult to numerically obtain the quark
propagator of the time like region which is need to
calculate meson masses.

Because of the difficulty for calculating hadron masses in
the covariant way we adopt instantaneous interactions which
include the confinement potential, the Coulomb potential
with the running coupling constant, and the 't~Hooft
determinant interaction \cite{thooft} as the intermediate
range interaction.  We show here that the 't~Hooft
determinant interaction makes the important contributions to
not only $\eta$ and $\eta^{\prime}$ meson's mass splitting
but other meson's masses, because it contributes to DCSB.

The outline of this paper is the following: In section 2, we
find the non-perturbative vacuum using Bogoliubov-Vallatin
transformation and write down the Salpeter equation which
describes meson states.  We also comment renormalization.
In section 3, we introduce the 't~Hooft determinant
interaction as a phenomenological interaction at
intermediate range and show our results.  Section4 is
devoted to summary.

\section{A phenomenological potential model
         for vacuum and meson spectra}
The Hamiltonian for quarks interacting through an
instantaneous fourth component potential is given as
follows:
\begin{eqnarray}
  H &=& H_{0} + H_{pot} \nonumber\\
    &=& \int\!d^{3}x \psi^{\dagger}(\Vec{x})
        (-i \Vec{\alpha} \cdot \nabla + \beta \hat{m})
        \psi(\Vec{x}) \nonumber\\
    & & \mbox{}
     +\;\frac{1}{2} \sum_{a} \int\!d^{3}x \int\!d^{3}y
      \left[ \psi^{\dagger}(\Vec{x}) \frac{\lambda^{a}}{2}
             \psi(\Vec{x}) \right]
      V(\Vec{x}-\Vec{y})
      \left[ \psi^{\dagger}(\Vec{y}) \frac{\lambda^{a}}{2}
             \psi(\Vec{y}) \right],
\label{eqn:hamiltonian}
\end{eqnarray}
where the row vector $\psi^{\dagger}$ for three flavour
quarks represents ($u^{\dagger}$ $d^{\dagger}$
$s^{\dagger}$), $\hat{m}$ is the current mass matrix:
\begin{equation}
  \hat{m} = {\rm diag}(m_{u}, m_{d}, m_{s}),
\end{equation}
and ${\displaystyle \frac{\lambda^{a}}{2}}$ are generators
of SU(3) colour group in the fundamental representation.

The potential is the sum of a linear colour-confinement
potential and a Coulombic potential with the running
coupling constant.  In momentum space it is written as
follows:
\begin{equation}
  V(q) = \frac{8 \pi \sigma}{q^{4}}
       + \frac{4 \pi \alpha(q^{2})}{q^{2}},
\end{equation}
where $\alpha(q^{2})$ is the running coupling constant which
varies with momentum transfer $q$.  According to
ref.\cite{weyers}, $\alpha(q^{2})$ is chosen such as
\begin{equation}
  \alpha (q^{2}) = \left\{
  \begin{array}{cl}
    {\displaystyle
     \frac{12 \pi}{(33-2N_{f}) \ln q^{2} / \Lambda_{QCD}^{2}}}
    & (q > q_{0}) \\
    & \\
    \alpha_{0} & (q \leq q_{0})
  \end{array}
  \right.,
\end{equation}
where
\begin{equation}
  q_{0} \equiv
  \Lambda_{QCD} \exp \frac{6\pi}{(33-2N_{f}) \alpha_{0}},
\end{equation}
and $N_{f}$ is a number of flavour.

In ordinary perturbation theory, quark fields are expanded
in terms of plain wave solutions.
\begin{equation}
  \psi(\Vec{x}) =
  \sum_{s} \int\!\!\frac{d^{3}p}{(2\pi)^{3/2}}
  \left[ a(\Vec{p}, s) u(\Vec{p}, s)
       + b^{\dagger}(-\Vec{p}, -s) v(-\Vec{p}, -s) \right]
  e^{i \Vec{p} \cdot \Vec{x}},
\end{equation}
where we suppress the colour, flavour and spinor indices.
The vacuum state $|0\rangle$ is defined as follows:
\begin{equation}
  a(\Vec{p},s) |0\rangle =
  b(\Vec{p},s) |0\rangle = 0.
\end{equation}
Since all other state vectors are constructed on this vacuum,
we call it perturbative vacuum.

To find the lower energy state than the perturbative vacuum,
we perform the Bogoliubov-Vallatin transformation, i.e.,
quark fields are expanded in terms of chiral transformed
spinors:
\begin{equation}
  \psi (\Vec{x}) =
  \sum_{s} \int\!\!\frac{d^{3}p}{(2\pi)^{3/2}}
  \left[ \alpha(\Vec{p}, s) U(\Vec{p}, s)
       + \beta^{\dagger}(-\Vec{p}, -s) V(\Vec{p}, s) \right]
  e^{i \Vec{p} \cdot \Vec{x}},
\end{equation}
where
\begin{eqnarray}
  U(\Vec{p}, s)
  & \equiv &
  \exp(\Vec{\gamma} \cdot \hat{\Vec{p}} \theta_{p})
  u(\Vec{p}, s), \\
  V(\Vec{p}, s)
  & \equiv &
  \exp(\Vec{\gamma} \cdot \hat{\Vec{p}} \theta_{p})
  v(-\Vec{p}, -s),
\end{eqnarray}
and $\theta_{p}$ is the chiral angle which shows amount of
chiral symmetry breaking.  $\alpha$ and $\beta$ are
annihilation operators of pseudo-particle and
pseudo-anti-particle, respectively.  These operators are
connected to those of particle as follows:
\begin{eqnarray}
  \alpha(\Vec{p},s) &=&
  \cos\theta_{p} \cdot a(\Vec{p},s)
  - \sin\theta_{p}
  \left(
    \begin{array}{cc}
      \Vec{\sigma} \cdot \hat{\Vec{p}} & 0 \\
      0 & \Vec{\sigma} \cdot \hat{\Vec{p}}
    \end{array}
  \right)
  \cdot b^{\dagger}(-\Vec{p},-s), \\
  \beta^{\dagger}(-\Vec{p},-s) &=&
  \sin\theta_{p}
  \left(
    \begin{array}{cc}
      \Vec{\sigma} \cdot \hat{\Vec{p}} & 0 \\
      0 & \Vec{\sigma} \cdot \hat{\Vec{p}}
    \end{array}
  \right)
  \cdot a(\Vec{p},s)
  + \cos\theta_{p} \cdot b^{\dagger}(-\Vec{p},-s).
\end{eqnarray}

Now, we define the non-perturbative vacuum state
$|0\rangle\!\rangle$ , which is the vacuum of
pseudo-particle and pseudo-anti-particle as
\begin{equation}
  \alpha(\Vec{p},s) |0\rangle\!\rangle =
  \beta (\Vec{p},s) |0\rangle\!\rangle = 0.
\end{equation}
It is useful to rewrite the Hamiltonian as the normal order
with respect to above operators.
\begin{equation}
  H = \varepsilon + :h: + :H_{4}: \;\;,
\end{equation}
$\varepsilon$ is the energy of non-perturbative vacuum:
\begin{eqnarray}
  \lefteqn{ \varepsilon = (-6) \int\!d^{3}p
    [ m \sin2(\theta_{p}+\delta_{p})
    + p \cos2(\theta_{p}+\delta_{p}) ]} \nonumber\\
  & & \mbox{}+
      \frac{2}{(2\pi)^{3}} \int\!d^{3}p\,d^{3}p'
      V(\Vec{p}-\Vec{p}') \nonumber\\
  & & \mbox{}-
      \frac{2}{(2\pi)^{3}} \int\!d^{3}p\,d^{3}p'
      V(\Vec{p}-\Vec{p}')
      \sin2(\theta_{p}+\delta_{p})
      \sin2(\theta_{p'}+\delta_{p'}) \nonumber\\
  & & \mbox{}-
      \frac{2}{(2\pi)^{3}} \int\!d^{3}p\,d^{3}p'
      V(\Vec{p}-\Vec{p}')  
      (\hat{\Vec{p}} \cdot \hat{\Vec{p}}')
      \cos2(\theta_{p}+\delta_{p})
      \cos2(\theta_{p'}+\delta_{p'}),
\end{eqnarray}
where the angle $\delta_{p}$ which leads to amounts of
explicit chiral symmetry breaking because of the current
mass is defined as
\begin{equation}
  \sin 2\delta_{p} \equiv \frac{m}{\sqrt{p^{2}+m^{2}}}.
\end{equation}
$h$ is the generalized one body Hamiltonian including self
energy effect:
\begin{eqnarray}
  \lefteqn{h = \int\!d^{3}x
    \psi^{\dagger}(\Vec{x})
    (-i \Vec{\alpha} \cdot \nabla + \beta m)
    \psi(\Vec{x})} \nonumber\\
  & & + \mbox{}
  \frac{4}{3} \times \frac{1}{2}
  \int d^{3}x d^{3}y \Int{p}
  V(\Vec{x}-\Vec{y}) e^{i \Vec{p} \cdot (\Vec{x}-\Vec{y})}
  \left[
    \psi^{\dagger}(\Vec{x})
    \{ 1 - 2 \Lambda_{-}(\Vec{p}) \}
    \psi(\Vec{y})
  \right],
\end{eqnarray}
where $\Lambda_{+}$ and $\Lambda_{-}$ are positive and
negative energy projection operators, respectively.

The angle $\theta_{p}$ is determined so that the energy of
non-perturbative vacuum may be lower than that of
perturbative vacuum.  Using variation principle, we get the
gap equation.
\begin{eqnarray}
  & &
  p \sin\phi_{p} - m \cos\phi_{p} \nonumber\\
  &+&
  \frac{2}{3} \sin\phi_{p}
  \Int{k} V(\Vec{p}-\Vec{k})
  \hat{\Vec{p}} \cdot \hat{\Vec{k}} \cos\phi_{k} \nonumber\\
  &-&
  \frac{2}{3} \cos\phi_{p}
  \Int{k} V(\Vec{p}-\Vec{k}) \sin\phi_{k} \nonumber\\
  &=& 0,
\end{eqnarray}
where $\phi_{p} \equiv 2(\theta_{p}+\delta_{p})$.

The procedure stated so far is independent of the particular
potential.  Since the gap equation with the Coulombic
potential contains divergent integral, we must renormalize
wavefunction and mass.  The wavefunction and mass
renormalization constants are given by
\begin{eqnarray}
  Z-1 &=&
  \frac{8}{9\pi}
  \int_{\textstyle q_{\scriptscriptstyle L}}dq \;
  q^{3} \frac{d}{dq^{2}}
  \left[
    \frac{\alpha(q^{2})}{q^{2}}
  \right], \\
  Z_{m}-1 &=&
  - \frac{4}{3\pi}
  \int_{\textstyle q_{\scriptscriptstyle L}}dq \;
  q \frac{\alpha(q^{2})}{q^{2}},
\end{eqnarray}
where we must introduce the lower cut-off $q_{L}$ to avoid
the logarithmic divergence.  Hence the renormalized gap
equation is
\begin{eqnarray}
  & &
  p\sin\phi_{p} + (Z-1)p\sin\phi_{p} \nonumber\\
  &-&
  m\cos\phi_{p} - (Z_{m}-1)m\cos\phi_{p} \nonumber\\
  &+&
  \frac{2}{3} \sin\phi_{p}
  \Int{k} V(\Vec{p}-\Vec{k})
  \hat{\Vec{p}} \cdot \hat{\Vec{k}} \cos\phi_{k} \nonumber\\
  &-&
  \frac{2}{3} \cos\phi_{p}
  \Int{k} V(\Vec{p}-\Vec{k}) \sin\phi_{k} \nonumber\\
  &=& 0.
\end{eqnarray}

If we solve this renormalized gap equation, we can calculate
the quark condensate.  However, even the quark condensate
subtracted the contribution of the explicit chiral symmetry
breaking is still divergent.  To compare with QCD sum rule
results for quark condensates, it is reasonable to introduce
the cut-off $\Lambda_{c}$=1GeV.  In this case the quark
condensate is calculated as follows,
\begin{equation}
  \langle \overline{\psi} \psi \rangle =
  3 \int^{\Lambda_{c}} \!\! \frac{d^{3}p}{(2\pi)^{3}}
  \left[
    \sin\phi_{p} - \frac{m}{\sqrt{p^{2}+m^{2}}}
  \right].
\end{equation}
We also make cut-off at 1GeV for the renormalized gap
equation.

Next step is to construct a meson state on the
non-perturbative vacuum.  We consider a meson state which is
a bound state of flavour $i$ quark and flavour $j$
anti-quark, and postulate that $i$ is not equal to $j$.  If
$i$ is equal to $j$, we must take the wave function of
isospin eigenstate.  To calculate meson masses, we have to
solve the Salpeter equation at zero center-of-momentum.  It
is known that the RPA equation is equivalent to the Salpeter
equation, but the RPA equations are derived for various type
of interactions more easily than the Salpeter equation.

We write down the RPA equations:
\begin{eqnarray}
  \langle\!\langle0| [ H,\hat{X} ] |m\rangle\!\rangle
    &=& (E_{0}-E_{m})
    \langle\!\langle0| \hat{X} |m\rangle\!\rangle, \\
  \langle\!\langle0| [ H,\hat{Y} ] |m\rangle\!\rangle
    &=& (E_{0}-E_{m})
    \langle\!\langle0| \hat{Y} |m\rangle\!\rangle,
\end{eqnarray}
where $|0\rangle\!\rangle$ and $|m\rangle\!\rangle$ are the
non-perturbative vacuum and the meson state, respectively.
Inserting the Hamiltonian Eq.(\ref{eqn:hamiltonian}) to
above equations, we obtain the following equations,
\begin{equation}
  [ E_{i}(k)+E_{j}(k) ] X(\Vec{k}) -
  \frac{4}{3} \Int{k^{\prime}} V(\Vec{k}-\Vec{k}^{\prime})
  [ X(\Vec{k}^{\prime})+Y(\Vec{k}^{\prime}) ]
  = M X(\Vec{k}),
\end{equation}
\begin{equation}
  [ E_{i}(k)+E_{j}(k) ] Y(\Vec{k}) -
  \frac{4}{3} \Int{k^{\prime}} V(\Vec{k}-\Vec{k}^{\prime})
  [ X(\Vec{k}^{\prime})+Y(\Vec{k}^{\prime}) ]
  = -M Y(\Vec{k}),
\end{equation}
where the forward and backward amplitudes of a meson state
are defined as
\begin{eqnarray}
  X(\Vec{k}) &=&
  \langle\!\langle0| \hat{X} |m\rangle\!\rangle =
  \sum_{a,S_{i},S_{j}}
  U^{i}(-\Vec{k}) \;
  \langle\!\langle0|
  \beta^{a}_{j}(\Vec{k}) \alpha^{a}_{i}(-\Vec{k})
  |m\rangle\!\rangle \;
  V^{j \dagger}(-\Vec{k}), \\
  Y(\Vec{k}) &=&
  \langle\!\langle0| \hat{Y} |m\rangle\!\rangle =
  \sum_{a,S_{i},S_{j}}
  V^{j}(-\Vec{k}) \;
  \langle\!\langle0|
  \alpha^{a \dagger}_{i}(-\Vec{k}) \beta^{a \dagger}_{j}(\Vec{k})
  |m\rangle\!\rangle \;
  U^{i \dagger}(-\Vec{k}),
\end{eqnarray}
$M$ is the mass of the meson state, and $E_{i}$ and $E_{j}$
are single particle energy of the particle and the
anti-particle, respectively.  In general above equations
depend on the center of mass momentum $\Vec{P}$ and the
relative momentum $\Vec{k}$.  Here we set $\Vec{P}=0$.
Dirac spinor $U(\Vec{k})$ and $V(\Vec{k})$ are eigen-vector
of the generalized one body Hamiltonian $h$ in momentum
space:
\begin{eqnarray}
  h(\Vec{k}) U(\Vec{k}) &=& E(\Vec{k}) U(\Vec{k}), \\
  h(\Vec{k}) V(\Vec{k}) &=& -E(\Vec{k}) V(\Vec{k}).
\end{eqnarray}
The forward and backward amplitudes are connected to the
Salpeter amplitude $\chi(\Vec{k})$ as follows:
\begin{eqnarray}
  \chi(\Vec{k})
  &=& \Lambda^{i}_{+}(\Vec{k})
      \chi(\Vec{k}) \Lambda^{j}_{-}(\Vec{k}) +
      \Lambda^{i}_{-}(\Vec{k})
      \chi(\Vec{k}) \Lambda^{j}_{+}(\Vec{k}) \nonumber\\
  &=& X(\Vec{k}) + Y(\Vec{k}),
\end{eqnarray}

Hence the RPA equations are converted to the Salpeter
equation:
\begin{eqnarray}
  &&
  [ E_{i}(k)+E_{j}(k) ]
  \left[ \Lambda^{i}_{+}(\Vec{k})\chi(\Vec{k})
  - \chi(\Vec{k})\Lambda^{j}_{+}(\Vec{k}) \right]
  \nonumber\\
  &-&
  \frac{4}{3} \int\!\!\frac{d^{3}k^{\prime}}{(2\pi)^{3}}
  V(\Vec{k}-\Vec{k}^{\prime})
  \left[ \Lambda^{i}_{+}(\Vec{k})\chi(\Vec{k}^{\prime})
  - \chi(\Vec{k}^{\prime})\Lambda^{j}_{+}(\Vec{k}) \right]
  = M \chi(\Vec{k}),
\label{eqn:salpeter}
\end{eqnarray}
and the Salpeter amplitude $\chi$ obeys the constraint:
\begin{equation}
  \Lambda^{i}_{+}(\Vec{k}) \chi(\Vec{k}) -
  \chi(\Vec{k}) \Lambda^{j}_{-}(\Vec{k}) = 0.
\label{eqn:constraint}
\end{equation}

To reduce the Salpeter equation, we expand the Salpeter
amplitude in terms of a complete set of Dirac matrices
\cite{yaouanc}
\begin{equation}
  \chi(\Vec{k})
  = \C{L}_{0}(\Vec{k})
  + \sum_{i=1}^{3} \C{L}_{i}(\Vec{k}) \rho_{i}
  + \vec{\C{N}}_{0}(\Vec{k}) \cdot \vec{\sigma}
  + \sum_{i=1}^{3} \vec{\C{N}}_{i}(\Vec{k})
    \cdot \rho_{i} \vec{\sigma}.
\end{equation}
The constraint Eq.(\ref{eqn:constraint}) demands
\begin{eqnarray}
  \C{L}_{0}(\Vec{k}) &=&
    i \frac{\sin\phi_{-}}{\cos\phi_{-}}
    \vec{\C{N}}_{2}(\Vec{k}) \cdot \Unit{k} , \\
  \C{L}_{3}(\Vec{k}) &=&
    - \frac{\cos\phi_{+}}{\sin\phi_{+}}
    \vec{\C{N}}_{1}(\Vec{k}) \cdot \Unit{k} , \\
  \vec{\C{N}}_{3}(\Vec{k}) &=&
    - \frac{\cos\phi_{+}}{\sin\phi_{+}} \C{L}_{1}(\Vec{k}) \Unit{k}
    -i \frac{\sin\phi_{-}}{\cos\phi_{-}}
    \vec{\C{N}}_{1}(\Vec{k}) \times \Unit{k} , \\
  \vec{\C{N}}_{0}(\Vec{k}) &=&
    i \frac{\sin\phi_{-}}{\cos\phi_{-}} \C{L}_{2}(\Vec{k}) \Unit{k}
    - \frac{\cos\phi_{+}}{\sin\phi_{+}}
    \vec{\C{N}}_{2}(\Vec{k}) \times \Unit{k} .
\end{eqnarray}
The general form of $\chi$ satisfying the constraint becomes
\begin{eqnarray}
  \chi (\Vec{k})
  &=&
  \C{L}_{1}
  \left[
    \rho_{1} - \frac{\cos\phi_{+}}{\sin\phi_{+}}
    \rho_{3} \vec{\sigma} \cdot \Unit{k}
  \right] \nonumber\\
  &+&
  \C{L}_{2}
  \left[
    \rho_{2} +i \frac{\sin\phi_{-}}{\cos\phi_{-}}
    \vec{\sigma} \cdot \Unit{k}
  \right] \nonumber\\
  &+&
  \vec{\C{N}}_{1} \cdot
  \left[
     \rho_{1} \vec{\sigma}
   - \frac{\cos\phi_{+}}{\sin\phi_{+}} \rho_{3} \Unit{k}
   +i\frac{\sin\phi_{-}}{\cos\phi_{-}}
     (\rho_{3} \vec{\sigma} \times \Unit{k})
  \right] \nonumber\\
  &+&
  \vec{\C{N}}_{2} \cdot
  \left[
     \rho_{2} \vec{\sigma}
   + \frac{\cos\phi_{+}}{\sin\phi_{+}}
     (\vec{\sigma} \times \Unit{k})
   +i\frac{\sin\phi_{-}}{\cos\phi_{-}} \Unit{k}
  \right].
\end{eqnarray}
The part of $\Lambda^{i}_{+}\chi^{\prime} -
\chi^{\prime}\Lambda^{j}_{+}$ in the left-hand side of
Eq.(\ref{eqn:salpeter}) is written as follows:
\begin{eqnarray}
  \lefteqn{
    \Lambda^{i}_{+}(\Vec{k}) \chi(\Vec{k}^{\prime})
  - \chi(\Vec{k}^{\prime}) \Lambda^{j}_{+}(\Vec{k})}
  \nonumber\\
  &=&
  \left\{
    -i\sin\phi_{+} \cos\phi_{-} \C{L}_{2}^{\prime}
    - \sin\phi_{+} \sin\phi_{-}
      (\vec{\C{N}}_{0}^{\prime} \cdot \Unit{k})
  \right\}
  \left[
    \rho_{1} - \frac{\cos\phi_{+}}{\sin\phi_{+}}
    \rho_{3} \vec{\sigma} \cdot \Unit{k}
  \right] \nonumber\\
  &+&
  \left\{
    i\sin\phi_{+}\cos\phi_{-} \C{L}_{1}^{\prime}
   -i\cos\phi_{+}\cos\phi_{-}
     (\vec{\C{N}}_{3}^{\prime} \cdot \Unit{k})
  \right\}
  \left[
    \rho_{2} +i \frac{\sin \phi_{-}}{\cos \phi_{-}}
    \vec{\sigma} \cdot \Unit{k}
  \right] \nonumber\\
  &+&
  \left\{
   -i\sin\phi_{+} \cos\phi_{-} \vec{\C{N}}_{2}^{\prime}
   - \sin\phi_{+} \sin\phi_{-} \C{L}_{0}^{\prime} \Unit{k}
   -i\cos\phi_{+} \cos\phi_{-}
      (\vec{\C{N}}_{0}^{\prime} \times \Unit{k})
  \right\} \nonumber\\
  && \mbox{} \cdot
  \left[
     \rho_{1} \vec{\sigma}
   - \frac{\cos\phi_{+}}{\sin\phi_{+}} \rho_{3} \Unit{k}
   +i\frac{\sin \phi_{-}}{\cos \phi_{-}}
     (\rho_{3} \vec{\sigma} \times \Unit{k})
  \right] \nonumber\\
  &+&
  \left\{
    i\sin\phi_{+} \cos\phi_{-} \vec{\C{N}}_{1}^{\prime} -
    i\cos\phi_{+} \cos\phi_{-} \C{L}_{3}^{\prime} \Unit{k} +
    \sin\phi_{+} \sin\phi_{-} (\vec{\C{N}}_{3}^{\prime} \times \Unit{k})
  \right\} \nonumber\\
  && \mbox{} \cdot
  \left[
     \rho_{2} \vec{\sigma}
   + \frac{\cos \phi_{+}}{\sin \phi_{+}}
     (\vec{\sigma} \times \Unit{k})
   +i\frac{\sin \phi_{-}}{\cos \phi_{-}} \Unit{k}
  \right],
\end{eqnarray}
where
\begin{equation}
  \phi_{\pm} \equiv \frac{\phi_{i} \pm \phi_{j}}{2},
\end{equation}
and the functions with the prime depend on
$\Vec{k}^{\prime}$.

Then the Salpeter equation is reduced to coupled equations,
\begin{eqnarray}
  && -\frac{E_{i}+E_{j}}{2}
    \frac{\sin\phi_{+}}{\cos\phi_{-}} \C{L}_{2} \nonumber\\
  &+& \frac{2}{3} \Int{k^{\prime}}
  V(\Vec{k}-\Vec{k}^{\prime})
  \left[
    \sin\phi_{+} \cos\phi_{-} \C{L}_{2}^{\prime}
   +\sin\phi_{+} \sin\phi_{-}
     \frac{\sin\phi_{-}^{\prime}}{\cos\phi_{-}^{\prime}}
     \C{L}_{2}^{\prime} \; \Unit{k}^{\prime} \cdot \Unit{k}
   \right.\nonumber\\
  &+& \left.
   i\sin\phi_{+} \sin\phi_{-}
     \frac{\cos\phi_{+}^{\prime}}{\sin\phi_{+}^{\prime}}
     (\vec{\C{N}}_{2}^{\prime} \times \Unit{k}^{\prime})
       \cdot \Unit{k}
  \right]
  = \frac{M}{2i} \C{L}_{1}, \\
&& \nonumber\\
&& \nonumber\\
  && \frac{E_{i}+E_{j}}{2}
    \frac{\cos\phi_{-}}{\sin\phi_{+}} \C{L}_{1} \nonumber\\
  &-& \frac{2}{3} \Int{k^{\prime}}
  V(\Vec{k}-\Vec{k}^{\prime})
  \left[
    \sin\phi_{+} \cos\phi_{-} \C{L}_{1}^{\prime}
   +\cos\phi_{+} \cos\phi_{-}
     \frac{\cos\phi_{+}^{\prime}}{\sin\phi_{+}^{\prime}}
     \C{L}_{1}^{\prime} \; \Unit{k}^{\prime} \cdot \Unit{k}
   \right.\nonumber\\
  &+& \left.
    i\cos\phi_{+} \cos\phi_{-}
      \frac{\sin\phi_{-}^{\prime}}{\cos\phi_{-}^{\prime}}
      (\vec{\C{N}}_{1}^{\prime} \times \Unit{k}^{\prime})
      \cdot \Unit{k}
  \right]
  = \frac{M}{2i} \C{L}_{2}, \\
&& \nonumber\\
&& \nonumber\\
  && \frac{E_{i}+E_{j}}{2}
    \left[
      -\frac{\cos\phi_{-}}{\sin\phi_{+}} \vec{\C{N}}_{2}
      +\left\{
        \frac{\cos^{2}\phi_{+} \cos\phi_{-}}{\sin\phi_{+}}
       -\frac{\sin\phi_{+} \sin^{2}\phi_{-}}{\cos\phi_{-}}
       \right\}
      (\vec{\C{N}}_{2} \cdot \Unit{k}) \Unit{k}
    \right] \nonumber\\
  &-& \frac{2}{3}
  \Int{k^{\prime}} V(\Vec{k}-\Vec{k}^{\prime})
  \left[
    -\sin\phi_{+} \cos\phi_{-} \vec{\C{N}}_{2}^{\prime}
    +\cos\phi_{+} \cos\phi_{-}
    \frac{\cos\phi_{+}^{\prime}}{\sin\phi_{+}^{\prime}}
    (\vec{\C{N}}_{2}^{\prime} \times \Unit{k}^{\prime})
    \times \Unit{k}
 \right.\nonumber\\
    &-& \left.
    \sin\phi_{+} \sin\phi_{-}
    \frac{\sin\phi_{-}^{\prime}}{\cos\phi_{-}^{\prime}}
    (\vec{\C{N}}_{2}^{\prime} \cdot \Unit{k}^{\prime}) \Unit{k}
    -i\cos\phi_{+} \cos\phi_{-}
    \frac{\sin\phi_{-}^{\prime}}{\cos\phi_{-}^{\prime}}
    \C{L}_{2}^{\prime} \Unit{k}^{\prime} \times \Unit{k}
  \right] = \frac{M}{2i} \vec{\C{N}}_{1}, \\
&& \nonumber\\
&& \nonumber\\
  && \frac{E_{i}+E_{j}}{2}
    \left[
      \frac{\sin\phi_{+}}{\cos\phi_{-}} \vec{\C{N}}_{1}
     +\left\{ \frac{\cos^{2}\phi_{+} \cos\phi_{-}}{\sin\phi_{+}}
     -\frac{\sin\phi_{+} \sin^{2}\phi_{-}}{\cos\phi_{-}} \right\}
       (\vec{\C{N}}_{1} \cdot \Unit{k}) \Unit{k}
     \right] \nonumber\\
  &-& \frac{2}{3}
  \Int{k^{\prime}} V(\Vec{k}-\Vec{k}^{\prime})
  \left[
    \sin\phi_{+} \cos\phi_{-} \vec{\C{N}}_{1}^{\prime}
    +\cos\phi_{+} \cos\phi_{-}
    \frac{\cos\phi_{+}^{\prime}}{\sin\phi_{+}^{\prime}}
    (\vec{\C{N}}_{1}^{\prime} \cdot \Unit{k}^{\prime}) \Unit{k}
 \right.\nonumber\\
    &-& \left.
    \sin\phi_{+} \sin\phi_{-}
    \frac{\sin\phi_{-}^{\prime}}{\cos\phi_{-}^{\prime}}
    (\vec{\C{N}}_{1}^{\prime} \times \Unit{k}^{\prime})
    \times \Unit{k}
    +i\sin\phi_{+} \sin\phi_{-}
    \frac{\cos\phi_{+}^{\prime}}{\sin\phi_{+}^{\prime}}
    \C{L}_{1}^{\prime} \Unit{k}^{\prime} \times \Unit{k}
    \right] = \frac{M}{2i} \vec{\C{N}}_{2}. \nonumber\\
\end{eqnarray}
If we make $i$ equal to $j$, we obtain the same equations as
those derived for one flavour quark by Yaouanc et,
al. \cite{yaouanc}

The pion decay constant $f_{\pi}$ is gotten by using the
pion wave function for the center of mass momentum
$\Vec{P}\ne0$ of pion according to the method of
ref.\cite{yaouanc} as follows:
\begin{equation}
  f_{\pi} =
  \left[
    3 \int^{\Lambda_{c}} \frac{d^{3}p}{(2\pi)^{3}}
    \frac{1}{M_{\pi}} \frac{2\C{L}_{1}}{\sin\phi_{p}}
  \right]^{1/2}.
\end{equation}
The Gell-Mann-Oakes-Renner relation \cite{gor} is given as
\begin{equation}
  -2 Z_{m} m_{u}
  \langle \overline{\psi}\psi \rangle_{0}
  = M_{\pi}^{2} f_{\pi}^{2},
\end{equation}
where $\langle\overline{\psi}\psi\rangle_{0}$ is the
condensate which is calculated by setting $m_{u}=0$.

Let us discuss numerical results.  Quark masses, strength of
linear potential, and $\Lambda_{QCD}$ are fixed as
$m_{u}=m_{d}=5$MeV, $m_{s}=190$MeV, $\sigma=4.69 {\rm
fm}^{-2}$, and $\Lambda_{QCD}=200$MeV/c.  The strength of
the Coulombic potential $\alpha_{0}$ and the lower cut-off
momentum $q_{L}$ included in the equations for $Z-1$ and
$Z_{m}-1$ are freely varied so as to fit to empirical values
of the pion decay constant, and condensation of $u$ and $s$
quarks.  For $\alpha_{0}=5.05$ and $q_{L}=0.86q_{0}$ we get
$f_{\pi}=93$MeV, $\langle \overline{u}u \rangle=-(263 {\rm
MeV})^{3}$ and $\langle \overline{s}s \rangle=-(240 {\rm
MeV})^{3}$.  Factors of counter terms, $Z-1$ and $Z_{m}-1$
are $-0.580$ and $-0.759$, respectively. Results of
calculated meson masses are shown in Table 2.  The pion mass
is about one half of the experimental value.  Also, other
meson masses are about 200MeV lower than experimental ones.
Single particle energies of quark and anti-quark are very
large because of the infrared properties of the linear
potential.  Actually, those are divergent at the momentum
$p=0$.  However, large parts of these single particle
energies are canceled out by the interaction between the
quark and the anti-quark.  In the present model the effect
of cancelation is too large.  Also, the absolute value of
factor $Z_{m}-1$ is too large, so pion and kaon masses
become much smaller.  The reason why the absolute value of
factor $Z_{m}-1$ is large is that we need to choose the
large coupling constant $\alpha_{0}$ of the Coulombic
interaction.

It is necessary to introduce such an intermediate range
interaction as does not essentially contribute the
quark-anti-quark interaction part for vector mesons although
it increases the dynamical mass of single quark.  The
absolute value of factor $Z_{m}-1$ reduces because the
coupling constant $\alpha_{0}$ of the Coulombic interaction
can be chosen to be small by introducing this interaction.
A candidate of this kind of interaction is the 't~Hooft
determinant interaction which mixes vacua of different
winding number.  So, let us add this interaction to the
Hamiltonian.

\section{The 't~Hooft determinant interaction}
As a phenomenological interaction of intermediate range, we
introduce the 't~Hooft determinant interaction:
\begin{equation}
  H_{det} =
  K \int d^{3}x
  \left\{
    \det \left[
      \overline{\psi}_{i}(\Vec{x})
      (1-\gamma_{5})
      \psi_{j}(\Vec{x})
    \right] +
    \det \left[
      \overline{\psi}_{i}(\Vec{x})
      (1+\gamma_{5})
      \psi_{j}(\Vec{x})
    \right]
  \right\},
\end{equation}
where the determinant is over flavour indices.

We apply the same procedure as previous section to the
Hamiltonian with the 't~Hooft determinant interaction:
\begin{equation}
  H = H_{0} + H_{pot} + H_{det}.
\end{equation}
The vacuum energy is
\begin{equation}
  \varepsilon =
  \sum_{f} \varepsilon_{f}
  + \varepsilon_{det},
\end{equation}
where $\varepsilon_{f}$ is flavour independent term and
\begin{equation}
  \varepsilon_{det} =
  -8N_{c}(2N_{c}^{2} + 3N_{c} + 1) K
  \Int{p} \sin\phi^{u}_{p}
  \Int{p} \sin\phi^{d}_{p}
  \Int{p} \sin\phi^{s}_{p}.
\end{equation}
Since the 't~Hooft determinant interaction mixes quark
flavours, the gap equation of $\theta^{u}_{p}$ and
$\theta^{s}_{p}$ becomes coupled equation:
\begin{eqnarray}
  & &
    p \sin\phi_{p} - m \cos\phi_{p}
    \nonumber\\
  &+&
    \frac{2}{3} \sin\phi_{p}
    \Int{k} V(\Vec{p}-\Vec{k})
    \hat{\Vec{p}} \cdot \hat{\Vec{k}} \cos\phi_{k}
    \nonumber\\
  &-&
    \frac{2}{3} \cos\phi_{p}
    \Int{k} V(\Vec{p}-\Vec{k}) \sin\phi_{k}
    \nonumber\\
  &-&
    8N_{c}(2N_{c}^{2}+3N_{c}+1) \cos\phi_{p}
    \Int{k} \sin\phi_{k} \Int{k} \sin\phi^{s}_{k}
    \nonumber\\
  &=& 0, \\
  & &
    p \sin\phi^{s}_{p} - m \cos\phi^{s}_{p}
    \nonumber\\
  &+&
    \frac{2}{3} \sin\phi^{s}_{p}
    \Int{k} V(\Vec{p}-\Vec{k})
    \hat{\Vec{p}} \cdot \hat{\Vec{k}} \cos\phi^{s}_{k}
    \nonumber\\
  &-&
    \frac{2}{3} \cos\phi^{s}_{p}
    \Int{k} V(\Vec{p}-\Vec{k}) \sin\phi^{s}_{k}
    \nonumber\\
  &-&
    8N_{c}(2N_{c}^{2}+3N_{c}+1) \cos\phi^{s}_{p}
    \left( \Int{k} \sin\phi_{k} \right)^{2}
    \nonumber\\
  &=& 0,
\end{eqnarray}
where
\begin{eqnarray}
  \phi_{p} &\equiv& 2(\theta^{u}_{p}+\delta^{u}_{p}), \\
  \phi^{s}_{p} &\equiv& 2(\theta^{s}_{p}+\delta^{s}_{p}),
\end{eqnarray}
and we set $m_{u}=m_{d}$, then the gap equation of
$\theta^{d}_{p}$ is the same equation as that of
$\theta^{u}_{p}$.

We adopt isospin eigenstates for meson states as follows:
\begin{eqnarray}
  \chi^{I=1} &=&
  \chi^{u\overline{d}}
\;\;,\;\;
  \frac{1}{\sqrt{2}}( \chi^{u\overline{u}} - \chi^{d\overline{d}} )
\;\;,\;\;
  \chi^{d\overline{u}} \\
  \chi^{I=0} &=&
  \frac{1}{\sqrt{2}}( \chi^{u\overline{u}} + \chi^{d\overline{d}} )
\;\;,\;\;
  \chi^{s\overline{s}}
\end{eqnarray}
The Salpeter equation for isospin $I=1$ and 0 states are
given by
\begin{eqnarray}
  &&
  \left[
    H(\Vec{k}),\chi^{I}(\Vec{k})
  \right]
  -
  \frac{4}{3}\Int{k'} V(\Vec{k}-\Vec{k}')
  \left[
    \Lambda_{+}(\Vec{k}),\chi^{I}(\Vec{k}')
  \right] \nonumber\\
  &+&(-)^{I+1}
  2(N_{c}+1)K \Int{p} \sin\phi^{s}_{p}
  \Int{k'}
  \sum_{i=\pm}
  \left[
    \Lambda_{+}(\Vec{k}),
    \gamma^{0}\Gamma_{i}
      \chi^{I}(\Vec{k}')
    \gamma^{0}\Gamma_{i}
  \right]
  \nonumber\\
  &+&(-)^{I+1}
  2N_{c}(N_{c}+1)K \Int{p} \sin\phi^{s}_{p}
  \Int{k'}
  \sum_{i=\pm}
  \Tr(\chi^{I}(\Vec{k}') \gamma^{0}\Gamma_{i})
  \left[ \Lambda_{+}(\Vec{k}),\gamma^{0}\Gamma_{i} \right]
  \nonumber\\
  &-&
  \delta_{I,0} \;
  2\sqrt{2}(N_{c}+1)K \Int{p} \sin\phi_{p}
  \Int{k'}
  \sum_{i=\pm}
  \left[
    \Lambda_{+}(\Vec{k}),
    \gamma^{0}\Gamma_{i}
      \chi^{s\overline{s}}(\Vec{k}')
    \gamma^{0}\Gamma_{i}
  \right]
  \nonumber\\
  &-&
  \delta_{I,0} \;
  2\sqrt{2}N_{c}(N_{c}+1)K \Int{p} \sin\phi_{p}
  \Int{k'}
  \sum_{i=\pm}
  \Tr(\chi^{s\overline{s}}(\Vec{k}') \gamma^{0}\Gamma_{i})
  \left[ \Lambda_{+}(\Vec{k}),\gamma^{0}\Gamma_{i} \right]
  \nonumber\\
  &=&
  M \chi^{I}(\Vec{k}),
\end{eqnarray}
\begin{eqnarray}
  &&
  \left[
    H^{s}(\Vec{k}),\chi^{s\overline{s}}(\Vec{k})
  \right]
  -
  \frac{4}{3}\Int{k'} V(\Vec{k}-\Vec{k}')
  \left[
    \Lambda^{s}_{+}(\Vec{k}),\chi^{s\overline{s}}(\Vec{k}')
  \right] \nonumber\\
  &-&
  \delta_{I,0} \;
  2\sqrt{2}(N_{c}+1)K \Int{p} \sin\phi_{p}
  \Int{k'}
  \sum_{i=\pm}
  \left[
    \Lambda^{s}_{+}(\Vec{k}),
    \gamma^{0}\Gamma_{i}
      \chi^{I}(\Vec{k}')
    \gamma^{0}\Gamma_{i}
  \right]
  \nonumber\\
  &-&
  \delta_{I,0} \;
  2\sqrt{2}N_{c}(N_{c}+1)K \Int{p} \sin\phi_{p}
  \Int{k'}
  \sum_{i=\pm}
  \Tr(\chi^{I}(\Vec{k}') \gamma^{0}\Gamma_{i})
  \left[ \Lambda^{s}_{+}(\Vec{k}),\gamma^{0}\Gamma_{i} \right]
  \nonumber\\
  &=&
  M \chi^{s\overline{s}}(\Vec{k}).
\end{eqnarray}
The Salpeter equation for isospin $I=1/2$ state is also
given by:
\begin{eqnarray}
  &&
  (E_{1}(\Vec{k}) + E_{2}(\Vec{k}))
  \left(
    \Lambda^{(1)}_{+}(\Vec{k}) \chi(\Vec{k}) -
    \chi(\Vec{k}) \Lambda^{(2)}_{+}(\Vec{k})
  \right)
  \nonumber\\
  &-&
  \frac{4}{3}\Int{k'} V(\Vec{k}-\Vec{k}')
  \left[
    \Lambda^{(1)}_{+}(\Vec{k}) \chi(\Vec{k}') -
    \chi(\Vec{k}') \Lambda^{(2)}_{+}(\Vec{k})
  \right] \nonumber\\
  &+&
  2(N_{c}+1)K \Int{p} \sin\phi^{(3)}_{p}
  \Int{k'}
  \sum_{i=\pm}
  \left[
    \Lambda^{(1)}_{+}(\Vec{k})
    \gamma^{0}\Gamma_{i}
      \chi(\Vec{k}')
    \gamma^{0}\Gamma_{i} -
    \gamma^{0}\Gamma_{i}
      \chi(\Vec{k}')
    \gamma^{0}\Gamma_{i}
    \Lambda^{(2)}_{+}(\Vec{k})
  \right]
  \nonumber\\
  &+&
  2N_{c}(N_{c}+1)K \Int{p} \sin\phi^{(3)}_{p}
  \Int{k'}
  \sum_{i=\pm}
  \Tr(\chi(\Vec{k}') \gamma^{0}\Gamma_{i})
  \left[
    \Lambda^{(1)}_{+}(\Vec{k})\gamma^{0}\Gamma_{i} -
    \gamma^{0}\Gamma_{i} \Lambda^{(2)}_{+}(\Vec{k})
  \right]
  \nonumber\\
  &=&
  M \chi(\Vec{k}),
\end{eqnarray}
where $\Gamma_{\pm} \equiv 1 \pm \gamma_{5}$.

The gap equation and the Salpeter equation including 't~
Hooft determinant interaction have to be solved.  We
introduce a kind of cut-off factor $e^{-p^{2}/p_{0}^{2}}$ to
the momentum integral parts in which the 't~Hooft
determinant interaction appears.  We choose the value of
parameter $p_{0}$ around 0.6GeV/c because the 't~Hooft
interaction is adopted as the intermediate interaction.  We
will show results of two cases of $p_{0}=0.6$GeV/c and
0.8GeV/c.  In each case, free parameters are the coupling
constant of the 't~Hooft determinant interaction $K$, the
coupling constant of the Coulombic interaction $\alpha_{0}$,
and the cut-off $p_{L}$.  Those are determined so as to fit
the pion decay constant and quark condensates.

Including the intermediate interaction, the dynamical mass
of quark increases.  Hence all meson masses increase and the
mass of kaon , especially, increases 150MeV.  However the
calculated pion mass is underestimated 30MeV to the
experimental value, since mass renormalization is more
effective than overestimate of quark condensate.

The results are summarized in table 1 and table 2.  For
meson masses, the difference between calculated values and
experimental values are within 70MeV.

\section{Summary}
In this paper, we constructed a phenomenological model which
reproduces the meson spectra.  At first, we considered the
phenomenological potential model which includes the
long-range linear potential and the short-range Coulombic
potential.  The meson masses were underestimated because the
effect of mass renormalization was too large due to the
large coupling constant of the Coulomb interaction which is
needed to reproduce $u$ and $s$ quark condensate and the
pion decay constant, and large parts of dynamical quark
masses were canceled out by the residual interaction energy.
Hence we introduced the 't~Hooft determinant interaction as
the intermediate interaction to increase the effective
dynamical mass of quarks without contributing to the
residual interaction of vector mesons.

The 't~Hooft determinant interaction played a important role
not to only $\eta,\eta^{\prime}$ mass difference, but other
meson masses through DCSB.  Setting parameters so as to fit
quark condensates and the pion decay constant, the
difference between calculated masses and experimental masses
could be within 70MeV.

Adding the gluon current to the quark current in the
Hamiltonian, according to ref.\cite{glueball}, we can
calculate the glueball spectra in the same method.

To describe hadron-hadron scattering processes in this
model, we need to describe moving meson.  However it is too
difficult to solve the Salpeter equation for non-zero center
of mass momentum without any approximation.  Hence we must
make an effort to find an appropriate approximation.

\newpage


\newpage
\begin{center}
  {\large\bf Table captions}
\end{center}

\begin{description}
  \item[Table 1] Quark condensates, the pion decay constant, 
and renormalization constants
  \item[Table 2] Meson masses
\end{description}

\newpage
\begin{center}
  {\bf Table 1} \\
  \vspace{3mm}
  \begin{tabular}{cccccc}
    \hline
    \hline
    & & & & & \\
         & $\langle \overline{u}u \rangle$ [${\rm MeV}^{3}$]
         & $\langle \overline{s}s \rangle$ [${\rm MeV}^{3}$]
         & $f_{\pi}$ [MeV]
         & $Z_{m}-1$
         & $Z-1$ \\
    & & & & & \\
    \hline
    \hline
    & & & & & \\
    Empirical & $-(230\pm25)^{3}$ & $-(200\pm30)^{3}$ & 93 \\
    & & & & & \\
    \hline
    $\alpha_{0}=5.05$ & & & & & \\
    $q_{L}=0.86q_{0}$ & $-(263)^{3}$ & $-(240)^{3}$ & 93 &
    -0.759 & -0.580 \\
    $K=0$ & & & & & \\
    \hline
    $\alpha_{0}=2.3$ & & & & & \\
    $q_{L}=0.8q_{0}$ & $-(283)^{3}$ & $-(219)^{3}$ & 92 &
    -0.605 & -0.344 \\
    $K=0.0485, p_{0}=4.0$ & & & & & \\
    \hline
    $\alpha_{0}=2.4$ & & & & & \\
    $q_{L}=0.84q_{0}$ & $-(282)^{3}$ & $-(227)^{3}$ & 92 &
    -0.568 & -0.329 \\
    $K=0.173, p_{0}=3.0$ & & & & & \\
    \hline
    \hline
  \end{tabular}
\end{center}

\vspace{1cm}

\begin{center}
  {\bf Table 2} \\
  \vspace{3mm}
  \begin{tabular}{ccccccccc}
    \hline
    \hline
    & & & & & & & & \\
         & $\pi$ & $\rho$ & $K$ & $K^{\star}$ & $\eta$ &
    $\eta^{\prime}$ & $\omega$ & $\phi$ \\
    & & & & & & & & \\
    \hline
    \hline
    & & & & & & & & \\
    Exp. & 140 & 770 & 498 & 892 & 547 &
    958 & 782 & 1020 \\
    & & & & & & & & \\
    \hline
    $\alpha_{0}=5.05$ & & & & & & & &  \\
    $q_{L}=0.86q_{0}$ &
    71 & 557 & 271 & 716 & & & & 823 \\
    $K=0$ & & & & & & & & \\
    \hline
    $\alpha_{0}=2.3$ & & & & & & & &  \\
    $q_{L}=0.8q_{0}$ & 103 & 769 & 420 & 867 & 473 &
    930 & 774 & 951 \\
    $K=0.0485, p_{0}=4.0$ & & & & & & & &  \\
    \hline
    $\alpha_{0}=2.4$ & & & & & & & &  \\
    $q_{L}=0.84q_{0}$ & 107 & 790 & 438 & 915 & 494 &
    1006 & 796 & 1017 \\
    $K=0.173, p_{0}=3.0$ & & & & & & & &  \\
    \hline
    \hline
  \end{tabular}
\end{center}

\end{document}